\documentclass{article}
\usepackage{ijcai22}

\usepackage{times}
\usepackage{soul}
\usepackage{url}
\usepackage[hidelinks]{hyperref}
\usepackage[utf8]{inputenc}
\usepackage[small]{caption}
\usepackage{graphicx}
\usepackage{amsmath}
\usepackage{amsthm}
\usepackage{booktabs}
\usepackage{cuted}
\usepackage{soul}

\usepackage[caption = false]{subfig}
\usepackage{amssymb}
\usepackage{xcolor}
\usepackage{multicol}
\usepackage{multirow}
\usepackage{makecell}
\usepackage{pifont}
\newcommand{\cmark}{\ding{51}}%
\newcommand{\xmark}{\ding{55}}%

\usepackage{wrapfig}
\usepackage{mathtools}
\usepackage{colortbl}
\definecolor{Gray}{gray}{0.85}
\newcolumntype{a}{>{\columncolor{Gray}}c}
\newcolumntype{x}{>{\columncolor{Gray}}l}

\definecolor{lightred}{RGB}{216,94,128}
\newcommand{\up}[1]{\scriptsize \textcolor{lightred}{${\blacktriangle#1\%}$}}
\newcommand{\upsymbol}{\textcolor{lightred}{${\blacktriangle}$}}
\definecolor{lightgreen}{RGB}{86,188,80}

\usepackage[ruled,titlenumbered,vlined]{algorithm2e}

\usepackage{diagbox}
\allowdisplaybreaks

\usepackage[capitalize]{cleveref}
\usepackage{xcolor}
\crefname{section}{Sec.}{Secs.}
\Crefname{section}{Section}{Sections}
\Crefname{table}{Table}{Tables}
\crefname{table}{Tab.}{Tabs.}
\Crefname{equation}{Equation}{Equations}
\crefname{equation}{Eq.}{Eqs.}
\Crefname{algorithm}{Algorithm}{Algorithms}
\crefname{algorithm}{Alg.}{Algs.}

\definecolor{airforceblue}{RGB}{149,169,216}
\definecolor{amber(sae/ece)}{RGB}{222,131,68}
\definecolor{amber}{RGB}{247,199,72}
\definecolor{darkgray}{RGB}{201,201,201}
\definecolor{darkred}{RGB}{210,132,129}

\newcommand{\DPST}{\textit{\textbf{DPST}}}

\title{\DPST : \textit{De Novo} Peptide Sequencing with Amino-Acid-Aware Transformers 
}
\author{}

\author{
Yan Yang$^{1,2}$\and
Zakir Hossain$^{1,2}$\and
Khandaker Asif$^2$ \and \\
Liyuan Pan$^{1,2}$ \and
Shafin Rahman$^3$ \and
Eric Stone$^{1,2}$ 
\affiliations
$^1$BDSI ANU, $^2$A\&F CSIRO, $^3$ECE NSU\\
\emails
\{u6169130\}@anu.edu.au
}

\begin{document}

\maketitle
\begin{abstract}



\textit{De novo} peptide sequencing aims to recover amino acid sequences of a peptide from tandem mass spectrometry (MS) data. Existing approaches for \textit{de novo} analysis enumerate MS evidence for all amino acid classes during inference. It leads to over-trimming on receptive fields of MS data and restricts MS evidence associated with following undecoded amino acids. Our approach, \DPST, circumvents these limitations with two key components: \textit{(1)} A \textit{confidence value aggregation} encoder to sketch spectrum representations according to amino-acid-based connectivity among MS; \textit{(2)} A \textit{global-local fusion} decoder to progressively assimilate contextualized spectrum representations with a predefined preconception of localized MS evidence and amino acid priors. Our components originate from a closed-form solution and selectively attend to informative amino-acid-aware MS representations. Through extensive empirical studies, we demonstrate the superiority of \DPST, showing that it outperforms state-of-the-art approaches by a margin of 12\% - 19\% peptide accuracy.

\end{abstract}

\section{Introduction}

Characterizing peptide sequences is the most challenging and attractive task in the proteomics study. Researchers usually sequence peptides from paired measurements of mass-to-charge ratio (m/z) and intensity for peptide fragments, generally called MS data. This can be done by comparison to a reference database, or by \textit{de novo} analysis in which such a database is not supplied.  The former can not generalize to unknown peptides. The latter has wider applications but needs to computationally infer assignments from MS data to each amino acid in a peptide. 

\begin{figure}[!t]
    \centering
    \begin{minipage}[c]{\linewidth}
    \includegraphics[width = \linewidth]{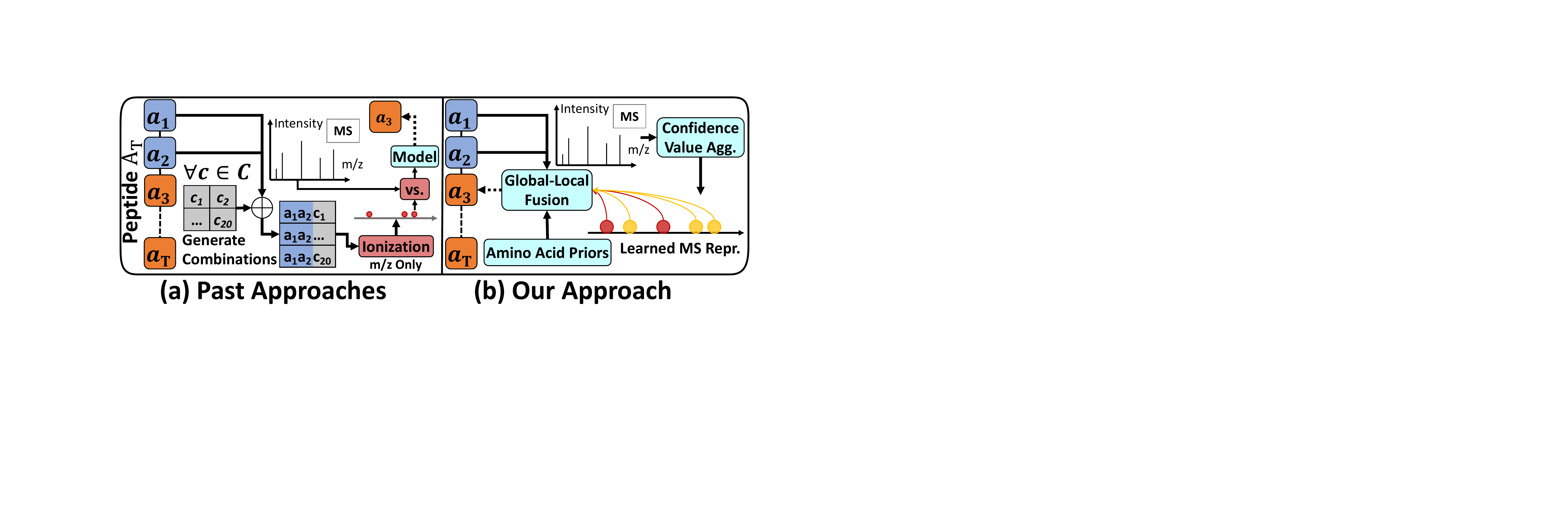}
    \vspace{-2em}
    \caption{\small
    Comparison with past approaches. The sequencing task aims to identify classes of amino acids from $a_1$ to $a_{\rm T}$ on a given peptide $\mathcal{A}_{\rm{T}}$,
    where $\rm T$ is the length of the peptide and $\mathcal{C}=\{c_1,\cdots,c_{20} \}$ is a set of amino acid classes. Here is an example of predicting an amino acid $a_3$.
    \textcolor{airforceblue}{Blue boxes} and \textcolor{amber(sae/ece)}{oranges boxes} indicate decoded and undecoded amino acids, respectively.
    \textcolor{gray}{Gray boxes} are amino acid classes.
    \textbf{(a)} Past approaches enumerate \textcolor{gray}{amino acid classes} to find possible evidence (\textcolor{darkred}{red ions}) for $a_3$. 
    \textbf{(b)} Our approach cracks the problem by introducing confidence value aggregation, amino acid priors,  as well as context (\textcolor{amber}{amber ions})
    and evidence (\textcolor{darkred}{red ions}) MS of \textcolor{airforceblue}{decoded amino acids}. (Best viewed in colour.)}
    \label{fig:intro}
    \end{minipage}
\end{figure}

\textit{De novo} peptide sequencing aims to identify amino acid sequences of a peptide $\mathcal{A}_{\rm{T}} = [a_{1}, a_{2}, \cdots, a_{\rm{T}}]$, where $a_i \in \mathcal{C}$.~Here, $\rm{T}$ is the peptide length and $\mathcal{C}$ is the amino acid classes set. Existing methods \cite{deepdenovo,deepdenovo2,smsnet} track MS for all possible amino acid combinations when predicting every next amino acid
(See \cref{fig:intro}. To predict $a_3$, they calculate m/z for every combination from $\{a_{1}, a_{2}, c_i \mid c_i \in \mathcal{C}\}$ and compare them with raw MS).

There exist several limitations: 
\textit{\textbf{(1)} Lack of MS reasoning.} 
Existing models only consider adjacent undecoded amino acids to perceive related MS. However, all spectrum data affords valuable information for synergistically peptide reasoning. These approaches only use completed MS for their LSTM hidden state initialization that pays barely to each inference step. Moreover, their methods prevent contextualized inference capability. If an amino acid produces a zero/low-intensity peak, they can hardly identify this amino acid from peaks associated with its previous and subsequent. Note, each peak contains a pair of m/z and intensity in MS data.
\textit{\textbf{(2)} Large search space.} Considering a standard amino acids set $\mathcal{C}=\{c_1,\cdots,c_{20} \}$ and $\lvert \mathcal{C} \rvert = 20$ \cite{standard_amino_acids}, the search space size of peptide sequencing is $20^{\rm{T}}$, with a given peptide at length $\rm T$. 
To the best of our knowledge, there are no existing methods to tackle this difficultly. We are the first to sharpen the problem's search space by supplying inductive bias to constrain feasible solutions in learning.

This paper studies a closed-form integer linear programming (ILP) \cite{ilp} solution to alleviate above mentioned obstacles in \textit{de novo} peptide sequencing.  As a result, we present our \DPST~framework with two key components (\cref{fig:intro} (b) describes our crucial idea). For \textit{MS reasoning}, we propose a confidence value aggregation encoder (\cref{sec:cvg}) to allow interactions among peaks originated from every same ion type. This formulation enables the learned spectrum representation to be aware of matched amino acids. It progressively constructs peptide-level evidence as encoder depth increases. In addition, it facilitates informative knowledge communication in the way of shrinking \textit{90\% problem search space}.  Furthermore, we propose a global-local fusion decoder (\cref{sec:glf}) to estimate undecoded amino acids under local and global MS clues. It aligns decoded amino acids with discriminative spectrum representation while bolstering from context evidence. We also derive amino acid priors for tuning received information from the encoder.


In this way,  our \DPST~preserves feasible context information during inferences. It concentrates on fulfilling a peptide prediction from shared MS representations instead of constructing diverse evidence for predicting amino acids. 
Hence, \DPST~achieves a larger improvement in peptide recall (\cref{fig:pes_acc}) than position-wise amino acid accuracy (\cref{fig:aa_acc}) when comparing with previous approaches.

Our main contribution are summarised as follow,
\begin{itemize}
    \item We develop a \DPST~framework based on transformer encoder-decoder architectures for \textit{de novo} peptide sequencing. 
    \item {We derive an ILP solution that serves as a theoretical underpinning for our confidence value aggregation encoder and global-local fusion decoder.}
    \item Our model outperforms state-of-the-art methods by 12\% - 19\% in peptide accuracy in extensive experiments.
\end{itemize}


\section{Related Work}
\noindent\textbf{\textit{De novo} Peptide Sequencing.} Recent advancements 
\cite{deepdenovo,deepdenovo2,smsnet,denovo1,denovo2,peaks} have shown potential to identify peptide sequences from MS data without including any reference databases. Among them, deep learning-based methods are most attractive due to their efficiencies in decoding and relaxations of a domain expert in an algorithm design. However, existing works rely on enumerating MS evidence for all amino acid classes at each decoding step.
The strategy inevitably suffers from bad quality MS data and harm model generality. To alleviate these problems, DeepNovo \cite{deepdenovo} and PointNovo \cite{deepdenovo2} employ dynamic algorithm-based post-processing steps to ensure predicted peptides are aligned with target peptide mass. SMSNet \cite{smsnet} additionally includes a database search process to substitute amino acids with low confidence scores. However, effort should still be inclined with learning. This paper aims to enhance model capability when encountering undesirable quality of MS data.


\noindent\textbf{Transformers.} A standard transformer paradigm \cite{transformer} is initially introduced for neural machine translation. 
Comparing with LSTM and CNN, it manipulates information with a broader horizon by attention mechanisms. Its simplicity and efficiencies have revolutionized extensive computing problems including image recognition \cite{vit}, object detection \cite{sotr,detr}, point cloud completion \cite{ptr,ptrpc}, and so on. Due to a quadratic complexity of a transformer and GPU memory bottlenecks, it has limited applicability in long sequences. To overcome the problem, Bigbird \cite{bigbird} and Performer \cite{performer} are introduced with linear approximation strategies for attention mechanisms. 
Nevertheless, a full-attention mechanism is not always dominating, and risks of introducing excessive noise when huge irrelevant features exist \cite{ptr} such as point cloud (i.e. MS)-based study. MS and peptides are similar to point clouds and sequences, respectively. Overall, our study explores a set (MS) to sequence (peptides) translation problem.


\begin{figure*}[!t]
    \centering
    \includegraphics[width=1.0\linewidth]{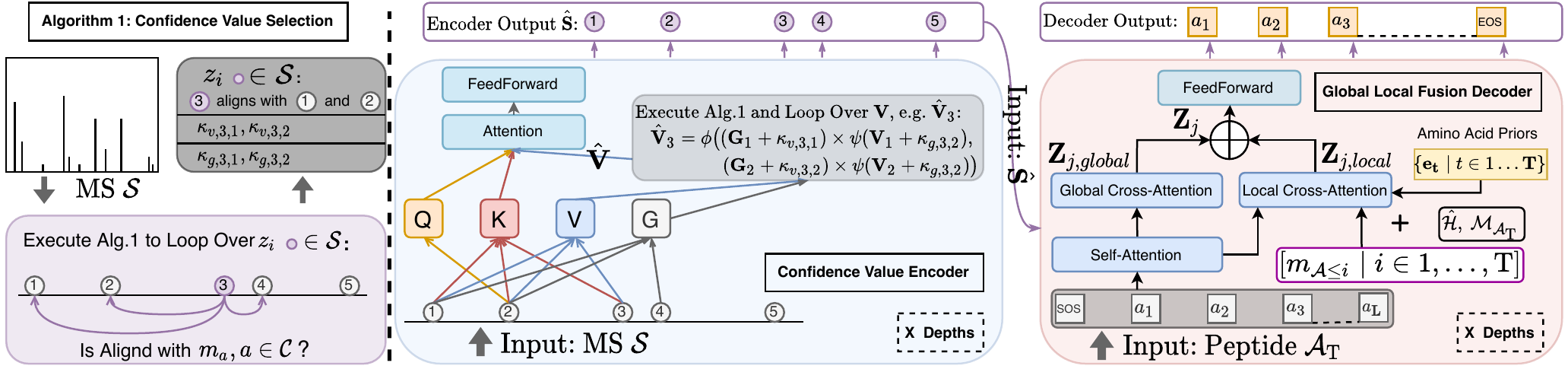}
    \vspace{-2em}
    \caption{
    Our network architecture. First, we execute \cref{alg:vgf} to retrieve strong connected pairwise peaks and obtain their distance. 
    Then, in the encoder, spectrum value representations depend on the number and quality of alignment among peaks.
    E.g., if the algorithm suggests $z_{1}$ and $z_{2}$ are aligned with $z_{3}$,  we aggregate a new value for $z_{3}$ named $\hat{\mathbf{V}}_{3}$ in accordance of their alignments/connectivity. Finally, in the decoder, the global branch attends to all spectrum representation $\hat{\mathbf{S}}$. The local branch merges with a localized spectrum representation of which peaks are produced by the latest decoded peptide fractions and is chosen by an empirical ionization function $\hat{\mathcal{H}}$. $\mathbf{e}_{t}$ offers prior knowledge of amino acids on a subsequent position to the decoder that helps to capture feasible solutions.}
    \label{fig:arch}
\end{figure*}

\section{Method}
We formulate our problem and analyze \DPST~design motivations in \cref{sec:sts}. Then, we elaborate encoder (Confidence Value Encoder) and decoder (Global Local Fusion Decoder) of \DPST~in \cref{sec:cvg} and \cref{sec:glf}, respectively. Our \DPST~framework is shown in \cref{fig:arch}.
\subsection{Overview}
\label{sec:sts}
\noindent \textbf{Problem Formulation.} Suppose we have a pair of MS $\mathcal{S}$ and peptide $\mathcal{A_{\rm{T}}}$. We define MS as an unordered point set $\mathcal{S} = \{s_{i} = (z_{i}, u_{i}) \in \mathbb{R}^2 \mid i = 1, \ldots, \rm{N} \}$, where $\rm{N}$, $z_{i}$ and $u_{i}$ are number of points, m/z and intensity, respectively.  
MS also has a machine-measured peptide mass 
{\small $m_{\mathcal{A}_{\rm{T}}} = \sum_{j = 1}^{\rm{T}} m_{a_{j}}$} as meta-information, where
$m_{a_{t}}$ is the mass of an amino acid $a_{t}$. Let {$\mathcal{A}_{\leq t} = [a_{1}, a_{2}, \cdots, a_{t}]$}. Our goal is to learn a mapping $\mathbf{F}(\cdot,\cdot)$ from {$\mathcal{A}_{\leq t}$} and $\mathcal{S}$ to $a_{t+1}$. We decompose $\mathbf{F}(\cdot,\cdot)$ into an encoder $\mathbf{E}(\cdot)$ and a decoder $\mathbf{D}(\cdot,\cdot)$ model by

\vspace*{-2mm}
{\footnotesize
\begin{equation}
    \mathbf{F}(\mathcal{A}_{\leq t},\mathcal{S}) = \mathbf{D}(\mathcal{A}_{\leq t}, \mathbf{E}(\mathcal{S})).
\end{equation}}

\noindent \textbf{Motivation.} Past approaches \cite{deepdenovo,deepdenovo2,smsnet} apply a CNN to understand possible MS evidence for each amino acid and connect them sequentially by an LSTM.  With over-trimmed MS evidence and a uni-direction nature of LSTM, they inevitably harm inferring from context. Though a vanilla transformer appears to resolve the above difficulties by modeling full MS together, it is not appropriate to our problem.  


Specifically, transformers apply multi-head attention to interact intra/inter-modal information flow and build their dependency. When tackling point cloud inputs, multi-head attention usually restricts memory throughput by interacting among k-Nearest Neighbors (kNN) for each point instance. Otherwise, the enormous number of points leads to GPU memory overload.
With an assumption of near points cohesion, these works further facilitate information flow by progressively downsampling and aggregating local features \cite{ptr,dgcnn}. 
%
However, such operations are not adequate in our task. First, the number of peaks/points increases with the peptide length, which leads to extra effort in managing MS size. Second, MS is sparsely distributed, and neighbourhood points are not likely to cooperatively form a local feature. This property differentiates MS data from image/video data. Moreover, a transformer usually results in low convergence speed or is stuck at a local minimal because of lacking appropriate inductive biases \cite{codetr} when expecting inter-modality outputs. ~~~~~~~~~~~~~~~~~~ ~~~ 

\begin{wrapfigure}{R}{0.3\linewidth}
    \includegraphics[width=1.0\linewidth, height = 0.9\linewidth]{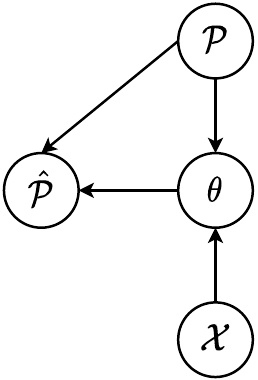}
  \vspace{-2em}
  \caption{\small Bayesian Network}
  \label{fig:bayes}
  \vspace{-1em}
\end{wrapfigure}
\indent By contrast, we leave out downsampling layers and compensate for information interactions by applying our confidence value aggregation strategy. It allows communication among every same type of ions, which revises spectrum representations according to amino acids between peak pairs. Next, we expect to learn ionization properties to connect amino acids with MS. Instead of modulating attention matrixes to speed up convergence \cite{codetr}, we align decoded amino acids with MS (Local Branch) while supplying amino acid priors to mitigate this phenomenon. With benefiting from our confidence value aggregation strategy, local branches can acquire and filter information about undecoded amino acids from aligned MS representations.

\noindent \textbf{ILP Solution.} We start by refactoring the \textit{de novo} task with an encoder and a decoder in a Bayesian perspective as

%
\vspace*{-2mm}
{\footnotesize
\begin{align}
    \Pr(a_{t+1} \mid \mathcal{A}_{\leq t}, \mathcal{S}, \theta) = \underbrace{\Pr(a_{t+1} \mid \mathcal{A}_{\leq t}, \hat{\mathbf{S}}, \theta_{D})}_{\text{Decoder}}  \underbrace{\Pr(\hat{\mathbf{S}} \mid \mathcal{S},\theta_{E})}_{\text{Encoder}} \ ,
    \nonumber 
\end{align}
}
where $\theta_E$ and $\theta_{D}$ are model parameters. $\hat{\mathbf{S}}$ is an intermediate MS representations. By applying the law of total probability, we introduce a class of discrete inductive bias hypothesis $\mathcal{X}$ while assuming a Bayesian network dependence as shown in \cref{fig:bayes}. 
Here, we use an empirical inductive bias hypothesis $\hat{x}$ for model training and implicitly assume the Dirac delta property for hypothesis classes.  Following \cite{uai}, we quantize Occam's razor with negative Kolmogorov complexity $km$ power of 2 in \cref{eq:km}. It encourages shortest, aka simple, inductive hypothesis descriptions. The `Model Learning' term can be measured by model loss objectives, while its second part describes how inductive bias affects learning model parameters.
To find the simple inductive bias hypothesis that best fits data and shrink the model parameter search space, we define

\vspace*{-2mm}
{\footnotesize
\begin{align}
    \log \Pr(&\hat{\mathbf{S}} \mid \mathcal{S}, \theta_{E}) = \log \sum_{x \in \mathcal{X}} \Pr(\hat{\mathbf{S}} \mid \mathcal{S}, \theta_{E})  ~ \frac{\Pr(x) ~ \Pr(\theta \mid \mathcal{S}, x)}{\Pr(\theta \mid \mathcal{S})} \nonumber \\
    \approx &  \underbrace{\log{\Pr(\hat{\mathbf{S}} \mid \mathcal{S}, \theta_{E})} \overbrace{\Pr(\theta \mid \mathcal{S},\hat{x})}^{\text{Inductive Bias Injection}}}_{\text{Model Learning}} - \underbrace{\log{\Pr(\theta \mid \mathcal{S})}}_{\text{Regularizers}} - \nonumber \\
    & \hspace{12em} \underbrace{Km(\hat{x}) \ .
    }_{\text{Complexity Penalty}} \label{eq:km}
\end{align}}

Then, we explore beneficial inductive bias in solving our problem with indications from an ILP of our problem. It also assists us in intuitively explaining our model constructions. 
Our objective is to find a peptide sequence with the least mass violation and satisfying relations with MS. Let $h(\cdot,\cdot) \in \mathcal{H}$ be an ionization function that takes mass, $m_{\mathcal{A}_{\leq t}}$ and $m_{\mathcal{A}_{\rm{T}}}$, of a peptide fraction $\mathcal{A}_{\leq t}$ and a completed peptide $\mathcal{A}_{\rm{T}}$ as inputs. Each $h(\cdot,\cdot)$ outputs a m/z for a distinct target ion.
For simplicity, we consider ionization functions \cite{deepdenovo2} that operate on N-terminus as an example  to define our ILP formulation. For each ionization function class $h(\cdot,\cdot)$ with a given peptide $\mathcal{A}_{\rm{T}}$, we have following ILP

\vspace*{-2mm}
{\footnotesize
\begin{align*}
    \text{min} & \quad \lvert \sum_{t \in 1 \ldots \rm{T}} \sum_{a \in \mathcal{C}}b_{t,a} \times m_{a} - m_{\mathcal{A}_{\rm{T}}} \rvert \\ 
    \text{s.t.}~~ & q_{i,t,a} - b_{t,a} = 0 & \forall_{i \in \rm{N}}, \forall_{t \in \rm{T}}, \forall_{a \in \mathcal{C}} \\
    &h(f(b,t) + m_{a}, m_{\mathcal{A}_{\rm{T}}}) = q_{i,t+1,a} \times z_{i} & \forall_{i \in \rm{N}}, \forall_{t \in \rm{T}}, \forall_{a \in \mathcal{C}} \\
    & \sum_{a \in \mathcal{C}} b_{t,a} = 1  & \forall_{t \in \rm{T}}\\
    & q_{i,t,a} \in \{0,1\}, b_{t,a} \in \{0,1\} & \forall_{i \in \rm{N}}, \forall_{t \in \rm{T}}, \forall_{a \in \mathcal{C}} \ ,
\end{align*}}
%
where $b_{[\cdot,\cdot]}$ and $q_{[\cdot,\cdot,\cdot]}$ are binary variables. Here, $b_{[\cdot,\cdot]}$ indicates selections of amino acid classes at each position, and $q_{[\cdot,\cdot,\cdot]}$ specifies the last amino acid class of an associating fragmented peptide for each peak (i.e. $\exists!_{i \in \rm{N}}~q_{i,2,a_{2}} = 1$ if the fragmented peptide is $a_{1}{a_{2}}$). We use 
{\small $f(b,t) = \sum_{t' = 0}^{t} \sum_{a \in \mathcal{C}} b_{t', a} \times m_{a}$} 
for calculating amino acid mass summation before its $t_{th}$ position. Note, we use $\rm{N}$ (number of peaks in MS) and $\rm{T}$ (length of a peptide) interchangeably to represent a scalar or a set of positive integers up to the scalar. 


The ILP above has a search space of $20^{\mathsf{T}}$ but we can utilize domain knowledge to shrink it. In MS, a majority of ions are 1$^+$ or~2$^+$ charged. Thus, regardless of ionization functions, we assume a pair of peaks $s_{i},s_{j}$ are connected if they satisfy  
\vspace*{-1mm}
{\footnotesize
\begin{equation}
    \lvert z_{i} - z_{j} \rvert \in \{m_{a} \mid a \in \mathcal{C}\} \cup \{\frac{1}{2} \times m_{a} \mid a \in \mathcal{C}\} \ . \label{eq:bridge}
\end{equation}}
%
Here, $\forall_{i,j \in \rm{N}^2}$ violates \cref{eq:bridge} at amino acid $a$, if $z_{i} < z_{j}$, $z_{j}$ is not associated with $a$ at a position after the associated amino acid of $z_{i}$, otherwise reverse relation exists. 
Once we identify $b_{[1,\cdot]}$ and $f_{[\cdot,1,\cdot]}$, 90\% infeasible amino acid selections can be excluded at an expense of quadratic time complexity under the strategy, as there are same mass at Leucine and Isoleucine.
We provide this constraint to our encoder during MS reasoning. Another possible approach \cite{deepdenovo} to exclude infeasible amino acids in $b_{[t+1,\cdot]}$ is considering mass constraints from decoded amino acids $b_{[\leq t,\cdot]}$ and the peptide $\mathcal{A}_{\rm{T}}$. However, this strategy only applied in the post-processing. 
In contrast, we extend the mass-based approach to quantify amino acid feasibility in the decoder during learning because it does not rely on MS that is usually incomplete. We can restrict our decoder to concentrate on feasible amino acids. In summary, with decoded amino acids and ionization functions, we can identify possible amino acids at the next position by a single-step computation. 


With the above analysis,  we derive a confidence value aggregation encoder, a global-local fusion decoder, and amino acid priors to respond to our problem.

\subsection{Confidence Value Aggregation Encoder}
\label{sec:cvg}
Our encoder uses a kNN-based multi-head self-attention \cite{ptr,ptrpc} as backbones and targets to learn informative spectrum representation. 


With \cref{eq:bridge}, we construct pairwise alignments for MS. The m/z $z_{i}$ of a peak $s_{i}$ (that aligns with other peaks of MS data) enables a description of an associated amino acid in focus.
In our formulation, spectrum representations $\hat{\mathbf{S}}$ hinge on alignment degrees among peaks. For each peak, $\delta$ is a hyper-parameter for selecting numbers of other peaks from same MS data to support itself, $\Phi(\cdot)$ records their index, and $\xi$ encodes distances between two peaks.  We also include an amino acid class (one-hot encoded) that bridges them in learning. See \cref{alg:vgf} for details.
This confidence value selection strategy, i.e. \cref{alg:vgf} in our kNN-based self-attention, allows the propagation of amino-acid-aware information flow among MS representations.
Each peak projects into a query $\mathbf{Q}$, key $\mathbf{K}$, value $\mathbf{V}$, and gate $\mathbf{G}$.  We revise $\mathbf{V}$ into $\hat{\mathbf{V}}$ considering reliance from others. 

\vspace*{-2mm}
{\footnotesize
\begin{align}
    &\mathbf{V} = \hat{\mathbf{S}} \mathbf{W}_{v}, \quad \quad \mathbf{G} = \hat{\mathbf{S}} \mathbf{W}_{g} \ , \\
    &\hat{\mathbf{V}}_{i} = \phi(\{(\mathbf{V}_{j} + \kappa_{v,i,j}) \times \psi(\mathbf{G}_{j} + \kappa_{g,i,j}) \mid j \in \Phi(i) \}) \ ,
    \label{eq:vg}
\end{align}}
%
where $\mathbf{W}_{v}$ and $\mathbf{W}_{g}$ are learnable weight matrix for $\mathbf{V}$ and $\mathbf{G}$, respectively. $\hat{\mathbf{V}}_{i}$ is a revised value representation for the peak $s_{i}$. Here,  $\phi$ is an aggregating function for unifying value representations, $\psi$ is a gating function such as Sigmoid,  $\mathbf{G}_{[\cdot]}$ filters value semantics, and  $\kappa_{[\cdot],i,j}$ are distance based representations between $i$ and $j$ that is used to calibrate value semantics. We then perform kNN-based self-attention on $\mathbf{Q}$, $\mathbf{K}$ and $\hat{\mathbf{V}}$. The value aggregating strategy facilitates kNN-based self-attention to reach consent by diagnosing relationships among local confident values.

\begin{algorithm}[!t]
\scriptsize{
\caption{Confidence Value Selection}\label{alg:vgf}
\SetKwInOut{Input}{Input}
\SetKwInOut{Output}{Output}
\Input{$\{z_{i} \mid i = 1,\ldots,\rm{N}\}$, $\delta$, $m_{a}$}
\tcc{$\delta$ specifies a number of peaks used to support each peak.}
\Output{$\{\Phi(i) \mid i = 1,\ldots,\rm{N} \}$, $\{\kappa_{v,i,j}\}$, $\{\kappa_{g,i,j}\}$}
\For{$i \in 1,\ldots,\rm{N}$}{
  $\Phi(i) = \{\}$ \;
  \For{$a \in \mathcal{C}$}{
   $\Phi(i)\text{.append}\big(\big\{ (\lvert d_{i,j} - r \times m_{a} \rvert, a) \mid j \in 1,\ldots \rm{N}, j \neq i, r \in \{\frac{1}{2}, 1\}\big\}\big)$, where $d_{i,j} = \lvert z_{i} - z_{j} \rvert$ \;
  }
  {$\Phi(i) = \text{Sort}(\Phi(i))[:\eta]$ \tcp*{Sort with  m/z difference.}}
  $\kappa_{v,i,j} = \text{FFN}(\xi(\Phi(i)_{j,0}), \text{OHE}(\Phi(i)_{j,1})) \quad \forall i,j \in 1,\ldots, \rm{N}$\;
  $\kappa_{g,i,j} = \text{FFN}(\xi(\Phi(i)_{j,0}), \text{OHE}(\Phi(i)_{j,1})) \quad \forall i,j \in 1,\ldots, \rm{N}$\;\tcc{$\xi$, OHE, FFN are a distance encoding function, an OneHotEncoder, and a feedforward neural network, respectively.}
  }
}
\end{algorithm}

\begin{figure*}[!t]
    \centering
    \includegraphics[width = 1.0\linewidth]{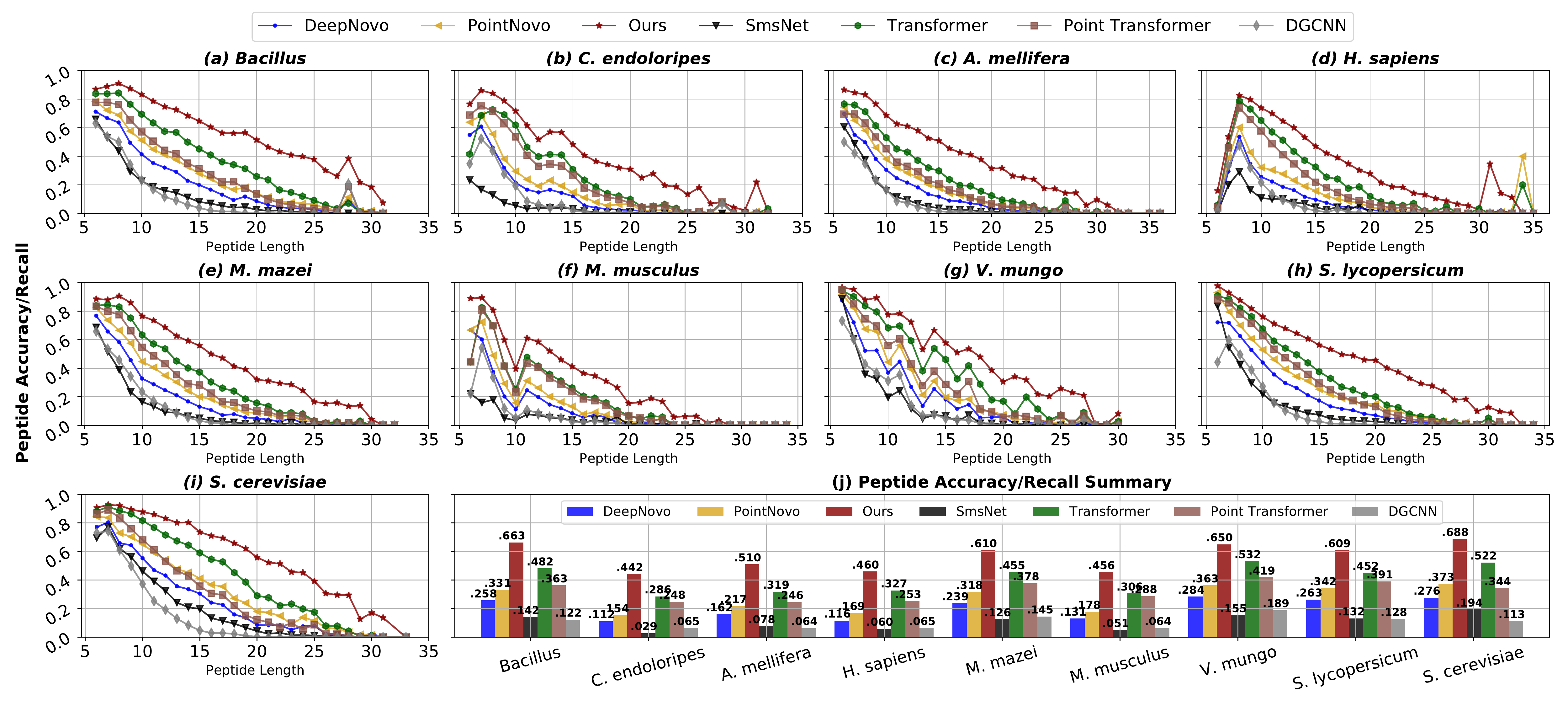}
    \vspace{-2.2em}
    \caption{\small Peptide accuracy/recall evaluation. (a)-(i) Model performance against peptide length on different datasets. (j) Overall model performances. (Best viewed in colour.)
     }
    \label{fig:pes_acc}
\end{figure*}
\begin{table*}[!t]
    \centering
        \vspace{-1em}
        \resizebox{1.0\linewidth}{!}{\begin{tabular}{llllllllll}
        \toprule
        Model & \textit{Bacillus} & \textit{C. endoloripes} & \textit{A. mellifera} & \textit{H. sapiens} & \textit{M. mazei} & \textit{M. musculus} & \textit{V. mungo} & \textit{S. lycopersicum} & \textit{S. cerevisiae} \\
        \midrule
        DeepNovo \cite{deepdenovo}&1.47\up{32}    &1.87\up{46}    &1.76\up{44}    &1.93\up{48}    &1.60\up{38}    &1.88\up{53}    &1.59\up{41}    &1.59\up{35}    &1.46\up{33}    \\ 
        PointNovo \cite{deepdenovo2}&1.30\up{17}    &1.66\up{30}    &1.54\up{26}    &1.70\up{31}    &1.42\up{22}    &1.66\up{35}    &1.41\up{25}    &1.42\up{20}    &1.28\up{16}    \\ 
        SMSNet  \cite{smsnet}&1.76\up{59}    &2.51\up{96}    &2.16\up{77}    &2.41\up{85}    &1.98\up{71}    &2.51\up{104}    &1.82\up{61}    &1.97\up{67}    &1.65\up{50}    \\ 
        Transformer$^{*}$ \cite{transformer}&1.33\up{20}    &1.89\up{48}    &1.71\up{40}    &1.75\up{35}    &1.47\up{27}    &1.66\up{35}    &1.33\up{18}    &1.50\up{27}    &1.31\up{19}    \\ 
        Point Transformer$^{*}$ \cite{ptr}&1.41\up{27}    &1.71\up{34}    &1.72\up{41}    &1.70\up{31}    &1.46\up{26}    &1.55\up{26}    &1.41\up{25}    &1.47\up{25}    &1.42\up{29}    \\ 
        DGCNN$^{*}$ \cite{dgcnn}&2.10\up{89}    &2.61\up{104}    &3.27\up{168}    &2.78\up{114}    &2.22\up{91}    &2.33\up{89}    &1.98\up{75}    &2.25\up{91}    &2.20\up{100}    \\ 
        \midrule
        \rowcolor{Gray}
        Ours & \textbf{1.11} & \textbf{1.28} & \textbf{1.22} & \textbf{1.30} & \textbf{1.16} & \textbf{1.23} & \textbf{1.13} & \textbf{1.18} & \textbf{1.10} \\
        \bottomrule
\end{tabular}}
\vspace{-1em}
\caption{\small Perplexity$\downarrow$ evaluations. `$\downarrow$' means the lower the better. We bold best results. $^{*}$ stands for minor changes that exist for adapting to our problem. \upsymbol~denotes deterioration ratio compared with our \DPST. See supplementary materials for details.}
\label{Table:eva}
\end{table*}

\subsection{Global-Local Fusion Decoder}
\label{sec:glf}
Our decoder targets to align MS representations $\hat{\mathbf{S}}$ to a peptide $\mathcal{A}$. A transformer decoder could catch ionization relations $\mathcal{H}$ and contextualized correspondences. However, transformers are general architectures for sequence problems. They are short of inductive bias about connectivity between amino acids and associated mass spectrum fractions. The aspect is reflected on low convergence speed or even stuck at local minimal during learning. 

We use two solutions to alleviate these problems. First, we speed up solving an ILP (\cref{sec:sts}) by considering the mass of decoded amino acids and the corresponding peptide. 
Here, models predict amino acid sequences in a fixed positional order. Given a prefix mass $m_{a_{\leq t}}$ and peptide mass $m_{\mathcal{A}_{\rm{T}}}$, we use dynamic programming to search the least mass violations for different classes in the first undecoded position \cite{deepdenovo}. Mathematically, we have

\vspace*{-2mm}
{\footnotesize
\begin{align}
   \mathbf{e}_{t} &= [\frac{1}{\sigma\sqrt{2\pi}} 
   \exp\left( -\frac{1}{2}\left(\frac{e_{a,t}}{\sigma}\right)^{\!2}\,\right)\mid a \in \mathcal{C}] \ ,\\ 
   e_{a,t} &= \min_{m_{U} \in \mathcal{C}^{+}} ~ \lvert m_{\mathcal{A}_{\rm{T}}} - m_{\mathbf{A} \leq t} - m_{a_{t+1}} - m_{\mathcal{U}} \rvert \ ,
\end{align}}
where $\mathbf{e}_{t}$ is a vector containing prior knowledge of amino acids at position $t+1$ that is formulated by a Gaussian distribution. {Note, $\mathcal{U} \in \mathcal{C}^{+}$ is an unknown variable and could be any amino acids combinations. We set a threshold for a mass of $\mathcal{U}$ in practice. With $\mathbf{e}_{t}$,  we can calibrate information obtained from MS $\hat{\mathbf{S}}$ after/before performing local attentions.}

Second, inspired by \cite{codetr}, we allure information bridging between representations of amino acid and MS. In \cref{sec:sts}, we describe reasoning following amino acids based on ionization functions, decoded amino acids and MS connectivity. Here, we assume a small set of ionization functions $\hat{\mathcal{H}}$. With a confidence value aggregation-equipped encoder, output spectrum representations are aware about the associated amino acid and its contexts. 

We can inherit information about the following undecoded amino acids when aligning a fraction of MS representations with each decoded amino acid through $\hat{h}(\cdot,\cdot) \in \hat{\mathcal{H}}$. Apart from full cross attentions 
{\footnotesize $\mathbf{Z}^{j}_{t,global}$}, 
we use a concurrent region-wise cross-attention layer {\footnotesize $\mathbf{Z}^{j}_{t,local}$} 
to directly inject the aforementioned strategy in learning. 
%
We perform cross attention for amino acids query $\mathbf{Q}$ with nearest MS key $\mathbf{K}$ and value $\mathbf{V}$ representations to capture spectrum information locally. Note, we calculate their distance based on ionization function outputs and m/z of peaks.
Let {\footnotesize $\mathbf{Z}^{j}_{t}$}
be a final output for $t_{th}$ amino acid in $j_{th}$ layer. We have

%
\vspace*{-2mm}
{\footnotesize
\begin{align}
    \mathbf{Z}^{j}_{t} &= \mathbf{Z}^{j}_{t,local} + \mathbf{Z}^{j}_{t,global} \ .
\end{align}}

\section{Experiments}
\noindent \textbf{Dataset.} We use DeepNovo dataset \cite{deepdenovo} to evaluate our model. The dataset contains nine distinct species. Following \cite{deepdenovo,deepdenovo2}, we remove peptide spectrum pairs that contain non-standard amino acids, MS with mass bigger than 3000 Da and peptides have a length over 50. Referring to supplementary materials for more details and experiments on other datasets.

\noindent \textbf{Methods for Comparisons.} We compare our proposed \DPST~with state-of-the-art methods including 
DeepNovo \cite{deepdenovo}, 
PointNovo \cite{deepdenovo2},
SMSNet \cite{smsnet},
Transformer$^*$ \cite{transformer},
Point Transformer$^*$ \cite{ptr} and
DGCNN$^*$ \cite{dgcnn}.  We train all models from scratch with recommended official settings except where indicated. Note, they did not release checkpoints for different folds, while we consider different amino acid classes compared with them. See supplementary material on how we adapt Transformer$^*$, Point Transformer$^*$ and DGCNN$^*$ to our task. Comparison with conventional \textit{de novo} sequencing algorithm is also provided in our supplementary material. 

\noindent \textbf{Implementation Details.} We train our model from scratch under Adam optimizer with a learning rate of $5 \times 10^{-4}$ for 20 epochs by using focal loss \cite{focalloss}. Our learning rate decays with cosine annealing schedule\cite{cos}.  \DPST~consists of 6 encoder layers and 6 decoder layers. The hidden dimension, feedforward layer expansion factor and the number of heads in attention are set to 128, 2, and 2, respectively. $\xi$ is an encoding function mixed with cosine and sine \cite{nerf}. See supplementary materials for details. Our code is available at 
\url{https://github.com/Yan98/DPST}.


\subsection{Results}
We use training, testing, and validation partitions from \cite{deepdenovo}. Eight of these folds are for training and validation, and the remaining fold is for testing. We present additional comparisons and analysis in supplementary materials.

\noindent \textbf{Perplexity Evaluation.} We present model perplexity across DeepNovo datasets in \cref{Table:eva}. A lower perplexity indicates a stronger model generalization ability. Our model outperforms the second-best model, PointNovo, by an absolute decrease from 0.18 to 0.43 on perplexity. During training, the performance of DeepNovo, PointNovo, and SMSNet have been saturated after several epochs. As they are over-reliant on the existence of corresponding MS evidence for each amino acid, their inferences are not echoing with contextualized knowledge (in contrast with our encoder). 
For Transformer, Point Transformer, and DGCNN, models are endued with context reasoning on mass spectrum data. However, they exile problem-specific inductive bias to aid learning.


\begin{figure}[!t]
    \centering
    \includegraphics[width = 1.0\linewidth]{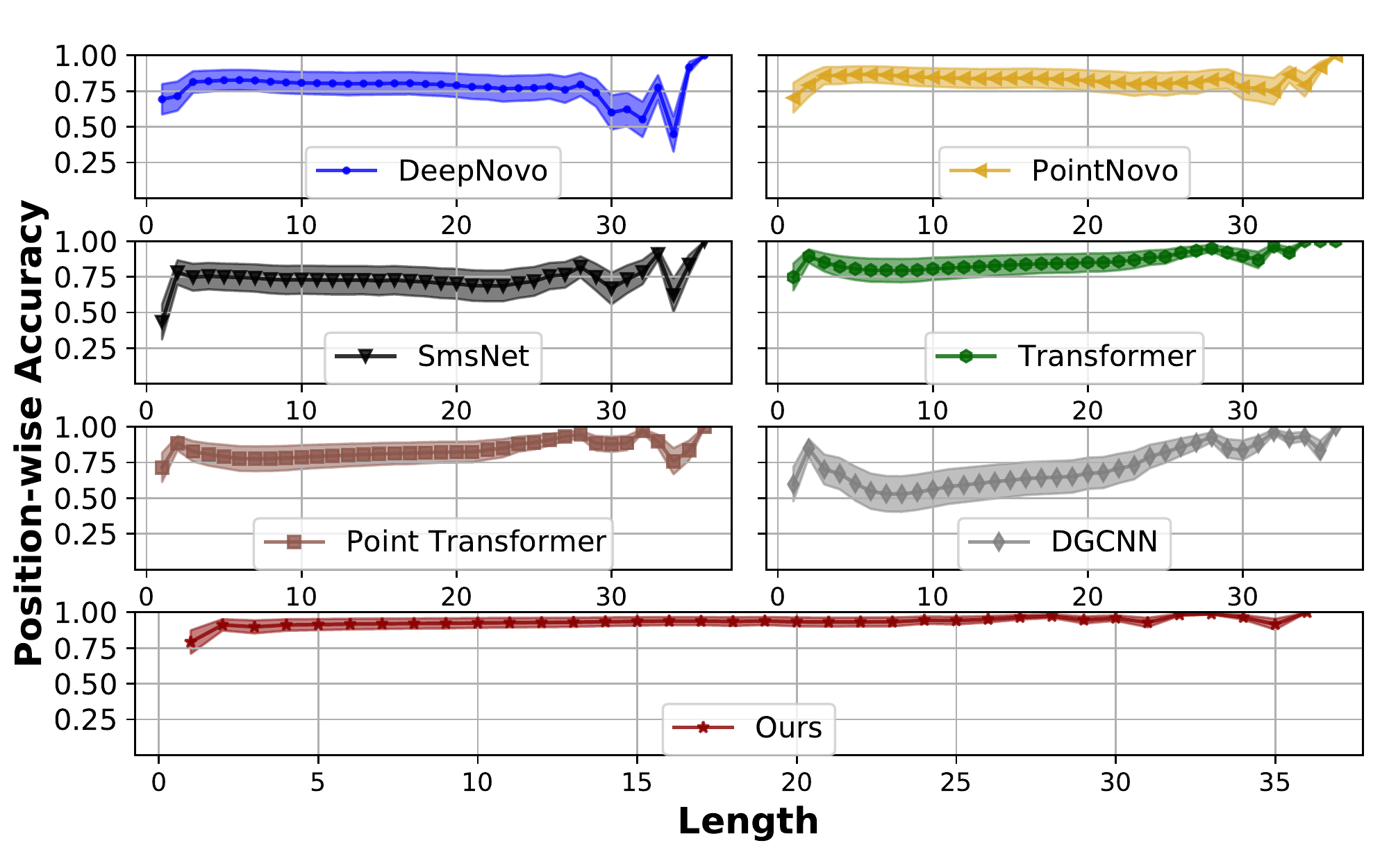}
    \vspace{-2.2em}
    \caption{Amino acid accuracy evaluation. Shadow area indicates the variance.}
    \label{fig:aa_acc}
\end{figure}
\noindent \textbf{Decoding Quality.} 
We evaluate peptide level accuracy (also known as peptide recall in literature) of our model against state-of-the-art methods by greedy decoding (see \cref{fig:pes_acc}) as the strategy directly reflects model capabilities. 
Our model consistently dominates other methods with peptide lengths from 5 to 35. In \cref{fig:pes_acc} (j), there is a margin of absolute accuracy improvement from 0.12 to 0.19 when compared with the second-best performing model. From \cref{fig:pes_acc} (a) to (i), all models lose decoding accuracies when peptide length increases - that is expected because the expectation of perfect decoding exponentially decreases with peptide length increments. We validate our claim by depicting position-wise (amino acid) accuracy in \cref{fig:aa_acc}, where our amino acid accuracy is more robust against peptide length variance compared with others. It is worth noting that  Transformer outperforms PointNovo in decoding quality but not in perplexity evaluation. It implies that enumerating all possible MS evidence for a first undecoded amino acid 
during inference is inadequate. Their models have weak context reasoning ability. Thus, they achieve wrong peptide decoding if MS evidence for any amino acid is absent.

\subsection{Ablation Study}
We perform the ablation study on the most challenging \textit{C.endoloripes} species in DeepNovo dataset \cite{deepdenovo,deepdenovo2,smsnet}. See supplementary material for additional ablation study. 

\noindent \textbf{Architecture Components.} We investigate components capability by conducting a detailed ablation study on \DPST~model architectures in \cref{Table:ablation}. Baseline model `A' is a vanilla transformer with kNN-based self-attention encoder. We ablate the transformer decoder (global branch) to local branches in model `B'. When using both branches (model `C'), the model perplexity has been slightly decreased. We then enable confidence value aggregation to model `A' and model `B'. The resulting models are named model `C' and model `D'.  Our confidence value aggregating strategy has improved model perplexity by 0.4 and 0.11. Our final model architecture (model `F') results in the best performance by successfully injecting inductive bias into the encoder (confidence value aggregation strategy) and decoder (local branch and amino acid priors). The number of local branches also has impacts on results as shown in \cref{fig:nkb}(a). The results suggest configuring local branches for the first four layers achieves the best results, which helps global branches to align with MS representation.

\begin{table}[!t]
    \centering
    \resizebox{\linewidth}{!}{\begin{tabular}{acccx}
    \toprule
     Model & Confidence Value Aggregation & Local Branch & Global Branch & Perplexity$\downarrow$ \\
     \toprule
     A & \xmark & \xmark & \cmark & 1.45\up{13}\\
     B & \xmark & \cmark & \xmark & 1.87\up{46}\\
     C & \xmark & \cmark & \cmark & 1.41\up{10}\\
     D & \cmark & \cmark & \xmark & 1.47\up{15}\\ 
     E & \cmark & \xmark & \cmark & 1.34\up{5}\\
     \midrule
     F & \cmark & \cmark & \cmark & \textbf{1.28}\\
    \bottomrule
    \end{tabular}}
    \vspace{-1em}
    \caption{Ablation study on model architectures. The \upsymbol~denotes deterioration ratio compared to the best architecture.
    }
    \label{Table:ablation}
\end{table}
\begin{table}[!t]
    \vspace{-2em}
    \captionsetup[subfloat]{labelfont=bf}
    \subfloat[][]{\renewcommand\arraystretch{0.5}
    {\begin{tabular}{a|c}
         \ Local Branch & Model Loss$\downarrow$  \\
         \midrule
         1 & 0.258\up{4.03} \\
         2 &  0.258\up{4.03} \\
         3 & 0.255\up{2.82}\\
         5 & 0.256\up{3.23}\\
         6 & 0.260\up{4.84}\\
         \midrule
         4 & \textbf{0.248} \\
         \bottomrule
    \end{tabular}}
    }
    \hspace{-0.5em}
    \captionsetup[subfloat]{labelfont=bf}
    \subfloat[][]{\renewcommand\arraystretch{0.5}
    {\begin{tabular}{aac}
         k & $\delta$ & Perplexity$\downarrow$  \\
         \midrule
         8 & 4 & 1.31\up{2.34}\\
         8 & 12 &  1.30\up{1.56} \\
         16 & 4 & 1.31\up{2.34}\\
         16 & 8 & 1.30\up{1.56}\\
         16 & 16 & 1.31\up{2.34}\\
         \midrule
         8  & 8  & \textbf{1.28}\\
         \bottomrule
    \end{tabular}}
    }
    \vspace{-1em}
    \caption{\textbf{(a)} Ablation on the number of local branches in the decoder. \textbf{(b)} We ablate on nearest neighbourhood k for self-attention and threshold $\delta$ for \cref{alg:vgf} in the encoder. The \upsymbol~denotes deterioration ratio compared with the best ablation.}
    \label{fig:nkb}
\end{table}

\noindent \textbf{Number of Neighbors.} 
We examine the number of nearest neighbors (k) used in self-attention of the encoder and the threshold ($\delta$) for selecting top confident values for \cref{alg:vgf} (see \cref{fig:nkb}~(b)). Setting k and $\delta$ both to 8 leads to best performance due to the following reasons. Most ions are $1^+$ or $2^+$ charged. Each peak can align with other peaks that have m/z smaller or bigger than itself. With diverse ionization functions, a bigger m/z does not imply an associated amino acid at a later position. The order of their associated amino acids is agnostic without knowing the types of ionization functions (which either operate on the N-terminus or C-terminus). Given an incomplete MS, we potentially benefit from the nearest neighbors of the desired peak. We can locate their m/z considering the distance of m/z between two other peaks. A larger or smaller value of k and $\delta$ would reduce the performance by introducing closer or farther $z_{i}$ in the mass spectrum. 



\section{Conclusions}
In this paper, we propose our \DPST~model by specializing the transformer architecture into the \textit{de novo} peptide sequencing task. Our \DPST~fuses a closed-form ILP solution into learning that adaptively concentrates on feasible and confident information flow during reasoning. Specifically, a value aggregation encoder obtains associated amino acids context information into MS representation from their pairwise spectrum alignments. Afterward, a global-local fusion decoder (with amino acids priors) inherit and filter such information for further inferences. Experimental results verify the superiority of our method in sequencing peptides from MS. Overall, the \DPST~prospects to serve as a new baseline method in the \textit{de novo} peptide sequencing. 

\bibliographystyle{named}
\bibliography{ijcai22_breif}

\end{document}